# Beat-spectrum design for 100-km-range optical correlation-domain reflectometry with localized 15-cm resolution

Takaki Kiyozumi[1,2*], Yuta Higa[1], Soshi Yoshida[2], Keisuke Motoda[1], Sze Yun Set[3], Shinji Yamashita[3,4], and Yosuke Mizuno[1,5]

[1]*Faculty of Engineering, Yokohama National University, Yokohama 240-8501, Japan*

[2]*School of Engineering, The University of Tokyo, Tokyo 153-8904, Japan*

[3]*Research Center for Advanced Science and Technology, The University of Tokyo, Tokyo 153-8904, Japan*

[4]*Graduate School of Engineering, The University of Tokyo, Tokyo 113-8656, Japan*

[5]*Institute for Multidisciplinary Sciences, Yokohama National University, Yokohama 240-8501, Japan*

*Author to whom correspondence should be addressed: kiyozumi-takaki@g.ecc.u-tokyo.ac.jp

**Conventional optical correlation-domain reflectometry (OCDR) based on sinusoidal frequency modulation exhibits a coupling between measurement range and spatial resolution because both are governed by the modulation frequency. Here, we formulate OCDR for arbitrary periodic frequency modulation and relate the modulation waveform to the resulting beat spectrum. By expressing the instantaneous optical frequency as a Fourier series, the beat spectrum is written as successive convolutions of the spectral contributions from the harmonic components. This formulation relates the harmonic composition of the modulation waveform to the spatial response. Periodic pseudo-random modulation (PPRM) was used to test this relation experimentally. We first measured the full-length reflectivity distribution along an approximately 100-km fiber using sinusoidal modulation and then performed PPRM-based random-access interrogation near the fiber end. In the local measurement, two closely spaced reflection points were resolved with a correlation-peak width of approximately 15 cm. These results show that beat-spectrum design can reduce the range–resolution coupling of conventional sinusoidal-modulation OCDR and combine long-range surveying with localized high-resolution interrogation.**

Optical reflectometry is a technique used to detect locate faults such as fiber breaks and high-loss events by launching light into a fiber under test (FUT) and analyzing the reflected signal[1]. As optical fiber networks continue to expand as critical communication infrastructure[2], the role of optical reflectometry has become increasingly important. In modern fiber networks, reflectometry is required not only to cover long-haul links on the order of 100 km[3], but also to resolve fine spatial features inside field-deployed components such as closures and connectors[4]. At the same time, practical deployment in large-scale networks requires a simple and cost-effective system configuration[5]. This trend has created a strong demand for optical reflectometry that simultaneously achieves long measurement range, decimeter-order spatial resolution, and low system complexity, requirements that are individually attainable but difficult to satisfy in combination.

Existing optical reflectometry techniques address these requirements only partially. Optical time-domain reflectometry (OTDR) identifies reflection locations by launching optical pulses into the FUT and measuring their round-trip time[6]. It offers a simple configuration and is well suited to long-distance measurements[3], but improving spatial resolution requires shortening the pulse width, which makes it difficult to achieve long measurement range and high spatial resolution simultaneously[7,8].

Optical frequency-domain reflectometry (OFDR) determines reflection locations by launching frequency-chirped light and analyzing the beat frequency between the incident and reflected signals[9,10]. OFDR can achieve micrometer-scale spatial resolution[11], but measurements over distances beyond the coherence length of the light source generally require compensation for laser phase noise or other additional techniques[12].

Optical correlation-domain reflectometry (OCDR) applies sinusoidal frequency modulation to the light source to generate a localized correlation peak that defines a measurement point on the FUT[13,14]. By scanning this peak along the fiber, the reflectivity distribution can be measured, and the measurement range can exceed the coherence length of the light source[15,16]. In addition, OCDR enables random-access interrogation at selected positions along the FUT using a simple interferometric configuration[17,18]. However, in conventional OCDR based on sinusoidal modulation[19,20,21], the measurement range and spatial resolution are governed by the modulation frequency, resulting in a range–resolution coupling that becomes restrictive in long-range measurements[22,23].

To overcome the above limitation of conventional OCDR, one approach is to extend the frequency modulation beyond a purely sinusoidal waveform and establish a general relation between the modulation waveform and the spatial response. Although various non-sinusoidal and noise-like modulation schemes have been proposed[24,25,26], a general formulation relating the spectral composition of an arbitrary periodic modulation waveform to the resulting beat spectrum and correlation-peak shape has not been established.

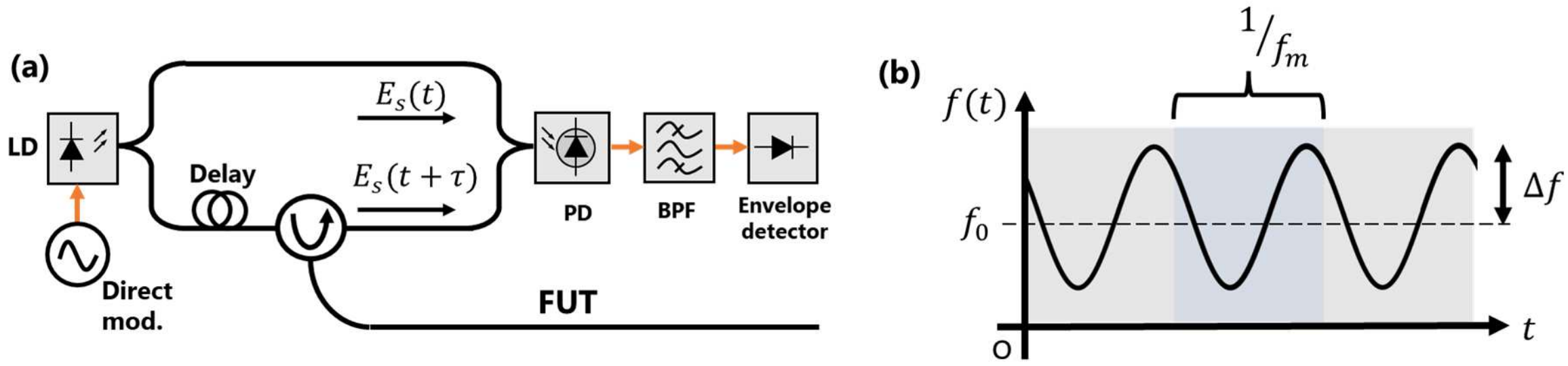

**FIG. 1. Fundamental scheme and modulation waveform for OCDR**. (a) Heterodyne interference detection in OCDR. A directly modulated laser diode (LD) output is split into a reference beam, $E(t)$, and a reflected beam, $E(t+\tau)$, which is delayed by $\tau$ after reflection in the fiber under test (FUT). The two beams interfere on a photodiode (PD), and the resulting beat signal is measured using heterodyne detection. The beat component is then isolated with a bandpass filter (BPF) centered at the beat frequency, and its amplitude is recovered via envelope detection. (b) Instantaneous optical frequency under sinusoidal frequency modulation. With respect to the center frequency $f_0$, a sinusoidal frequency sweep with amplitude $\Delta f$ is repeated with a period of $1/f_m$.

In this study, we establish a general formulation of OCDR under arbitrary periodic frequency modulation. By expressing the instantaneous optical frequency of the light source as a Fourier series, we show that the heterodyne interference signal can be decomposed into contributions from individual harmonic components. This formulation shows that the beat spectrum is synthesized through successive convolution of the spectral contributions from the harmonic components, relating the modulation waveform to the spatial response. Based on this relation, the spectral composition of the modulation waveform can be designed to sharpen the correlation peak and suppress sidelobes. As a proof of concept, we measure the reflection distribution along an approximately 100-km fiber and subsequently use periodic pseudo-random modulation for random-access interrogation near the fiber end, achieving a spatial resolution of 15 cm. These results demonstrate that beat-spectrum design can relax the range–resolution coupling of conventional sinusoidal-modulation OCDR.

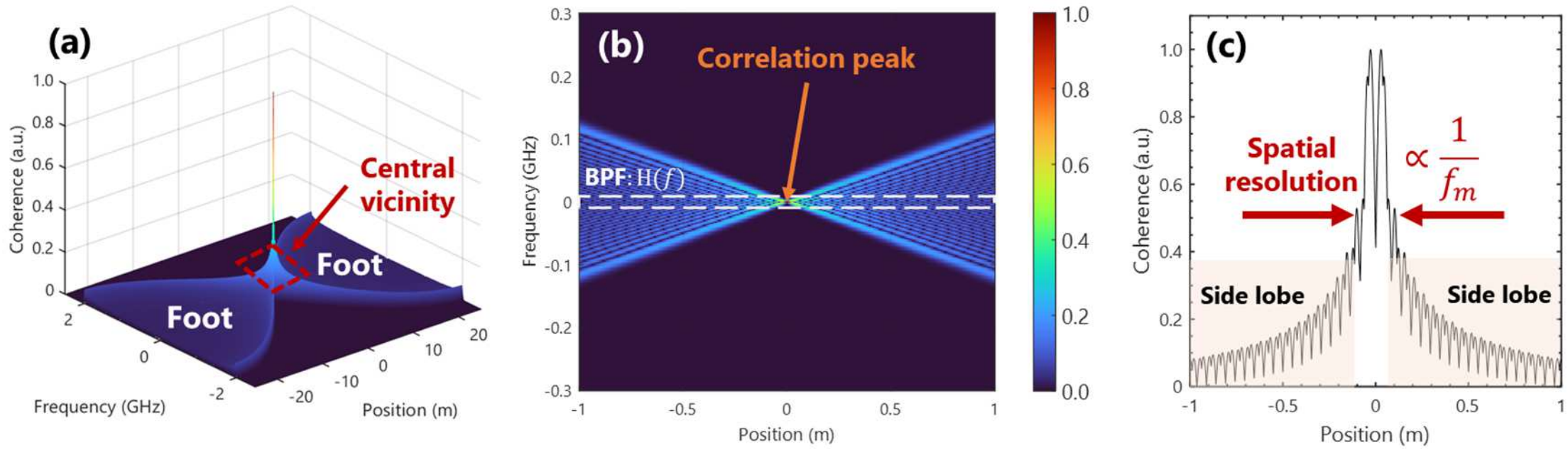

**FIG. 2. Magnitude of beat spectrum and correlation-peak shape under sinusoidal frequency modulation.** (a) Overview of the beat spectrum magnitude. A correlation peak appears at the center, with feet spreading outward. The modulation amplitude and frequency are $\Delta f = 1.0$ GHz and $f_m = 2$ MHz, respectively. (b) The central vicinity with a filter response $H(f)$ whose bandwidth is 9 MHz. The reflected-peak response is obtained by integrating the product of the beat spectrum and $H(f)$. (c) Reflection (correlation) peak profile. The response exhibits sidelobes originating from feet of the beat spectrum. The peak width is inversely proportional to the modulation frequency $f_m$.

# Principles of conventional OCDR

We first briefly review the operating principle of conventional OCDR based on sinusoidal frequency modulation to clarify the origin of the range–resolution trade-off considered in this study. Figure 1a schematically shows the basic configuration of an OCDR system. The light reflected from the FUT is interfered with a reference beam in an interferometric configuration, and a beat signal is obtained via heterodyne detection. Under sinusoidal frequency modulation, the complex optical field of the incident light can be written as

$$E_s(t) = \exp\left(j2\pi \int_0^t f(t')dt'\right), \tag{1}$$

where $\mathrm{j} = \sqrt{-1}$. The instantaneous optical frequency $f(t')$ is given by

$$f(t) = f_0 + \Delta f \sin(2\pi f_m t), \tag{2}$$

with $f_0$ the center optical frequency, $\Delta f$ the frequency-modulation amplitude, and $f_m$ the modulation frequency (Fig. 1b).

Let $\tau$ denote the relative time delay corresponding to the optical path difference between the reference arm and the signal reflected at a position $x$ from the zero optical path difference (ZOPD). The distance–delay relationship is

$$x = \frac{c\tau}{2n}\left(\Leftrightarrow \tau = \frac{2nx}{c}\right), \tag{3}$$

where $c$ is the speed of light in vacuum and $n$ is the refractive index (typically $n \simeq 1.46$ for silica fibers). The beat-spectrum component $S_A(x, f)$, obtained by Fourier transforming the interference term $E_s^*(t)\, E_s(t-\tau)$, consists of discrete spectral lines spaced by $f_m$ and can be expressed as

$$S_A(x, f) = \mathcal{F}[E_s^*(t) E_s(t-\tau)] = \sum_{\nu=-\infty}^{\infty} J_\nu\left(2\frac{\Delta f}{f_m}\sin\left(\frac{2\pi f_m n x}{c}\right)\right)\delta(f+\nu f_m), \tag{4}$$

where $J_\nu(\cdot)$is the $\nu$-th order Bessel function of the first kind, $\delta(\cdot)$ is the Dirac delta function, and $\mathcal{F}[\cdot]$ denotes the Fourier transform (Fig. 2a). For simplicity, the heterodyne frequency shift is omitted here, because it only translates the beat spectrum along the frequency axis without changing its envelope.

As indicated by Eq. (4), the beat spectrum is composed of discrete lines at $\nu f_m$, and the amplitude of each line varies with position $x$through the term $\sin(2\pi f_m n x/c)$. When the argument of this sine term becomes zero, the beat-spectrum power is concentrated in the central spectral line, corresponding to the strongest correlation. These positions function as measurement points and are referred to as correlation peaks. Taking the ZOPD as $x = 0$, correlation peaks appear at positions satisfying

$$\mathrm{x} = K\frac{c}{2nf_m}, \quad K = 0, \pm 1, \pm 2, \dots. \tag{5}$$

Here, $K$ denotes the correlation-peak order: the 0th peak is located at the ZOPD ($x = 0$), and higher-order peaks are periodically distributed along the FUT. By tuning $f_m$, the peak spacing can be varied, enabling scanning of the measurement point along the FUT. In addition, the position of the 0th correlation peak (ZOPD) can be adjusted by inserting a delay line in one arm of the interferometer. Because a selected nonzero-order correlation peak can be positioned along the FUT by tuning $f_m$, OCDR allows selective interrogation of a specific point on the FUT; this capability is commonly termed random access.

If multiple correlation peaks coexist within the FUT, however, it becomes difficult to attribute a measured reflection to a particular peak. Consequently, the unambiguous measurement range of OCDR is limited by the spacing between adjacent correlation peaks, yielding

$$\Delta x_{\mathrm{range}} = \frac{c}{2nf_m}. \tag{6}$$

Next, by filtering the beat spectrum using analog circuitry and/or digital signal processing, one obtains the spatial response (reflection-peak shape) along the fiber. Let $H(f)$ be the amplitude transfer function of the filter. The resulting peak shape $\gamma(x)$ is given by

$$|\gamma(x)| = \left| \int_{-\infty}^{\infty} S_A(x, f) \times H(f)\ df \right|. \tag{7}$$

The process described by Eq. (7) is illustrated in Fig. 2b and Fig. 2c. As shown in Fig. 2c, the shape of $\gamma(x)$ depends on the modulation frequency $f_m$. For fixed $f_z$ and $\Delta f$, a higher $f_m$ produces a narrower peak. For example, when a bandpass filter centered at $f_z$ with a bandwidth not exceeding $f_m$ is employed, the correlation-peak width (spatial resolution) is[19]

$$\Delta x \approx \frac{2 f_z c}{\pi n f_m \Delta f}. \tag{8}$$

Equations (6) and (8) indicate that, in conventional OCDR based on sinusoidal modulation, both the measurement range $\Delta x_{\mathrm{range}}$ and the reflection-peak width $\Delta x$ are governed by the modulation frequency $f_m$. This coupling gives rise to an intrinsic trade-off between range and spatial resolution[20,21], which limits the achievable resolution to the meter scale for fiber lengths of several tens of kilometers.

In addition, as shown in Fig. 2c, sinusoidal modulation generates sidelobes originating from the feet of the beat spectrum. These sidelobes appear as undesired components in the spatial response, reducing the ability to distinguish closely spaced reflections. The spectral feet can also produce spurious reflection peaks, known as ghost peaks, through spectral folding in heterodyne detection[27] (supplementary material, S1).

Therefore, effective suppression of sidelobes requires control of how the beat spectrum evolves with position. In particular, the beat spectrum should broaden rapidly in frequency as the position moves away from the correlation point, thereby reducing the spectral power remaining within the detection bandwidth.

## Generalized frequency modulation and beat-spectrum synthesis through harmonic convolution

To meet the design requirement identified above, it is necessary to move beyond purely sinusoidal frequency modulation and to consider more general modulation waveforms. In OCDR, the frequency modulation applied to the light source is periodic, which allows the instantaneous optical frequency to be expressed in a general form without loss of generality. This generalization provides a natural framework for analyzing how arbitrary periodic modulation waveforms determine the structure of the beat spectrum and, consequently, the spatial response of the system.

Because the frequency modulation applied to the light source is periodic, the instantaneous optical frequency $f_{\mathrm{arb}}(t)$ can be expressed as a Fourier series:

$$f_{\mathrm{arb}}(t) = f_0 + \sum_{k=1}^{\infty} \Delta f_k \sin(2\pi k f_m t + \alpha_k)\,. \tag{9}$$

Here, $\Delta f_k$, $k f_m$, and $\alpha_k$ represent the modulation amplitude, modulation frequency, and initial phase of the $k$-th harmonic component, respectively, while $f_m$ denotes the fundamental modulation frequency. As the optical phase is given by the time integral of the instantaneous frequency, the corresponding complex optical field can be written as

$$\begin{aligned} E_{\mathrm{arb}}(t) &= E_0 \exp\left\{ j2\pi \int_0^t f_{\mathrm{arb}}(t')dt' \right\} \\ &= E_0 \exp(j2\pi f_0 t) \exp\left\{ j2\pi \sum_{k=1}^{\infty} \int_0^t \Delta f_k \sin(2\pi k f_m t' + \alpha_k) dt' \right\} \\ &= E_0 \exp(j2\pi f_0 t) \prod_{k=1}^{\infty} \exp\left\{ j2\pi \int_0^t \Delta f_k \sin(2\pi k f_m t' + \alpha_k) dt' \right\}. \end{aligned} \tag{10}$$

When the generalized optical field given by Eq. (10) interferes with a delayed replica with a relative delay $\tau$, the heterodyne interference term can be expressed in a separable form with respect to the individual harmonic components:

$$\begin{aligned} I_{\mathrm{arb}}(t,\tau) &\propto E_{arb}(t+\tau)^* E_{arb}(t) \\ &= \prod_{k=1}^{\infty} \exp j2\pi \int_t^{t+\tau} \Delta f_k \sin(2\pi f_m t' + \alpha_k) dt'. \end{aligned} \tag{11}$$

This product form indicates that the phase contribution can be factorized into terms associated with the individual harmonic components. Using the convolution property of the Fourier transform, namely that multiplication in the time domain corresponds to convolution in the frequency domain, the beat spectrum $S_{\mathrm{arb}}(\tau, f)$ can be expressed as

$$S_{\mathrm{arb}}(\tau, f) = S_{(f_m, \Delta f_1)}(\tau, f) *_f S_{(2f_m, \Delta f_2)}(\tau, f) *_f \cdots *_f S_{(kf_m, \Delta f_k)}(\tau, f) *_f \cdots. \tag{12}$$

Here, $S_{(kf_m, \Delta f_k)}(\tau, f)$ denotes the beat spectrum generated by the $k$-th harmonic component alone, and $*_f$ represents convolution with respect to the frequency variable $f$. Notably, each $S_{(kf_m, \Delta f_k)}$ takes the same functional form as the conventional sinusoidal-modulation case (Eq. (4)), with the substitutions $f_m \rightarrow k f_m$ and $\Delta f \rightarrow \Delta f_k$.

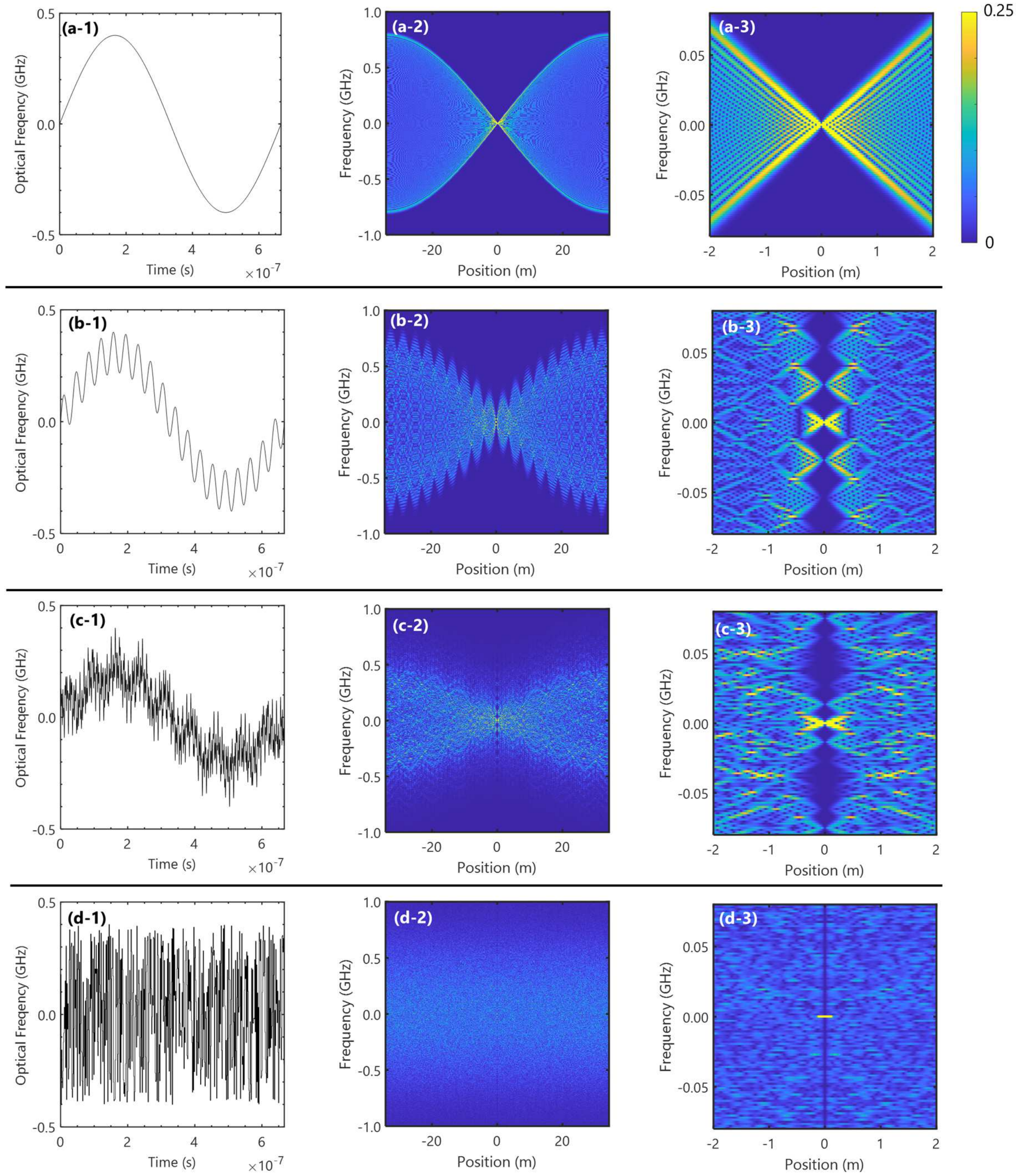

**FIG. 3. Beat-spectrum magnitude simulated based on Eq. (12)**. Each column shows (1) the modulation waveform, (2) an overview of the resulting beat spectrum, and (3) a magnified view around the spectral center. The color scale is normalized to a maximum value of 0.25 to clearly show the correlation peak. Each row corresponds to (a) sinusoidal modulation, $f_a(t) = \Delta f \sin(2\pi f_m t)$; (b) two-tone modulation, $f_b(t) = \Delta f\{\sin(2\pi f_m t) + 0.3\sin(2\pi\ 18 f_m t)\}$; (c) multi-harmonic modulation, $f_c(t) = \Delta f\{\sin(2\pi f_m t) + 0.3\sum_{k\in\{9,50,107,222\}} \sin(2\pi k f_m t)\}$; and (d) periodic pseudo-random modulation with a 500-point sequence of uniform random numbers $U(-\Delta f, \Delta f)$. The modulation parameters are $\Delta f = 0.4$ GHz and $f_m = 1.5$ MHz.

Equation (12) describes the beat spectrum as the successive convolution of the spectral contributions from the harmonic components. As additional harmonic components are introduced into the modulation waveform, the corresponding single-tone beat spectra are successively convolved along the frequency axis. As a result, the overall beat spectrum can be reshaped by controlling the amplitudes and phases of the harmonic components. This formulation makes it possible to control the spatial response of OCDR through the harmonic composition of the modulation waveform, rather than by adjusting a single modulation frequency.

The beat-spectrum synthesis process described by Eq. (12) is illustrated in Fig. 3, which compares simulated beat spectra for different periodic modulation waveforms with increasing numbers of harmonic components. In the case of conventional sinusoidal modulation [Fig. 3a], the beat spectrum exhibits pronounced spectral feet. As higher-order harmonic components are added [Figs. 3b and 3c], the beat spectrum evolves toward a more distributed profile, in which the spectral energy is spread over a wider frequency range. For PPRM, which contains many harmonic components [Fig. 3d], the spectral components are distributed over a still wider frequency range, producing a more localized response around the correlation point. The simulated spectra are consistent with the convolution description in Eq. (12).

# Experimental methods

## Experimental setup

To experimentally verify the beat-spectrum design principle established in the preceding sections, we constructed the OCDR system shown in Fig. 4. A laser diode operating at a center wavelength of approximately 1550 nm with a linewidth of 20 MHz was used as the light source. The output light was split by an optical coupler into a reference arm and a probe arm connected to the FUT. To adjust the position of the ZOPD, a 200-km fiber delay line was inserted in the probe arm.

The FUT connected to this system is shown in Fig. 5. It consisted of fiber spools with lengths of approximately 19.5 km, 25 km, and 50 km connected in series. Angled physical contact (APC) and physical contact (PC) connectors were intentionally introduced to create discrete reflection points labeled A, B, and C. In addition, a ~25 cm patch cable was attached at the end of the FUT to generate closely spaced reflection points C and D. This configuration provides well-defined reflection points near the fiber end, enabling quantitative evaluation of

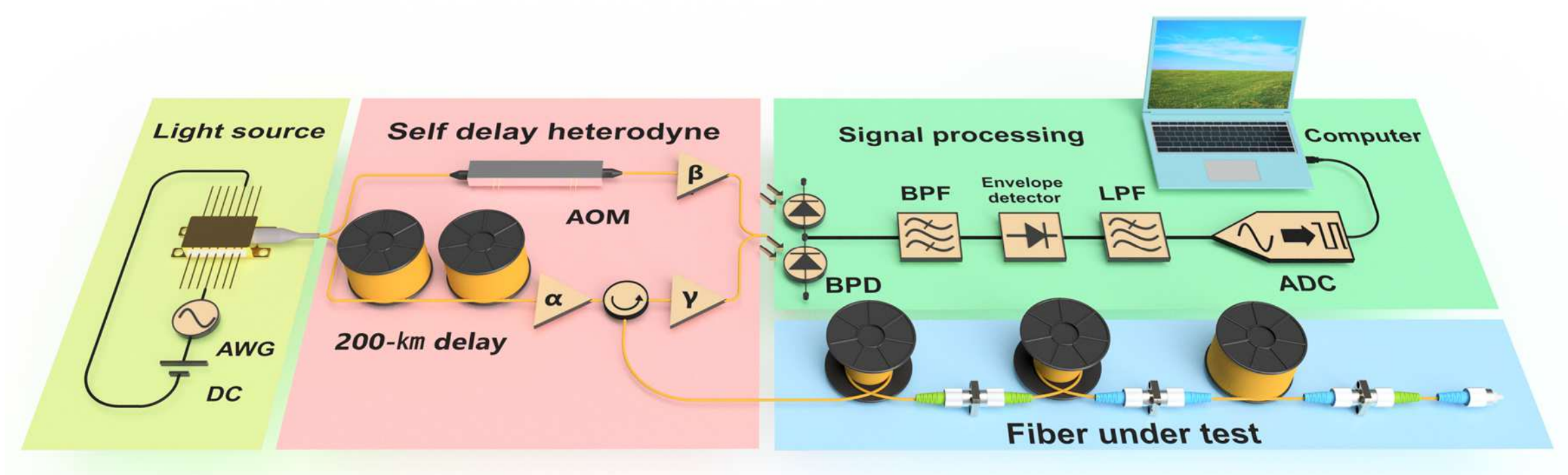

**FIG. 4.** Schematic of the OCDR experimental system, consisting of a self-delay heterodyne detection unit, the fiber under test, and the signal-processing chain (AWG: arbitrary waveform current generator; DC: direct-current source; BPD: balanced photodetector; BPF: band-pass filter; EDFA: erbium-doped fiber amplifier; ADC: analog-to-digital converter).

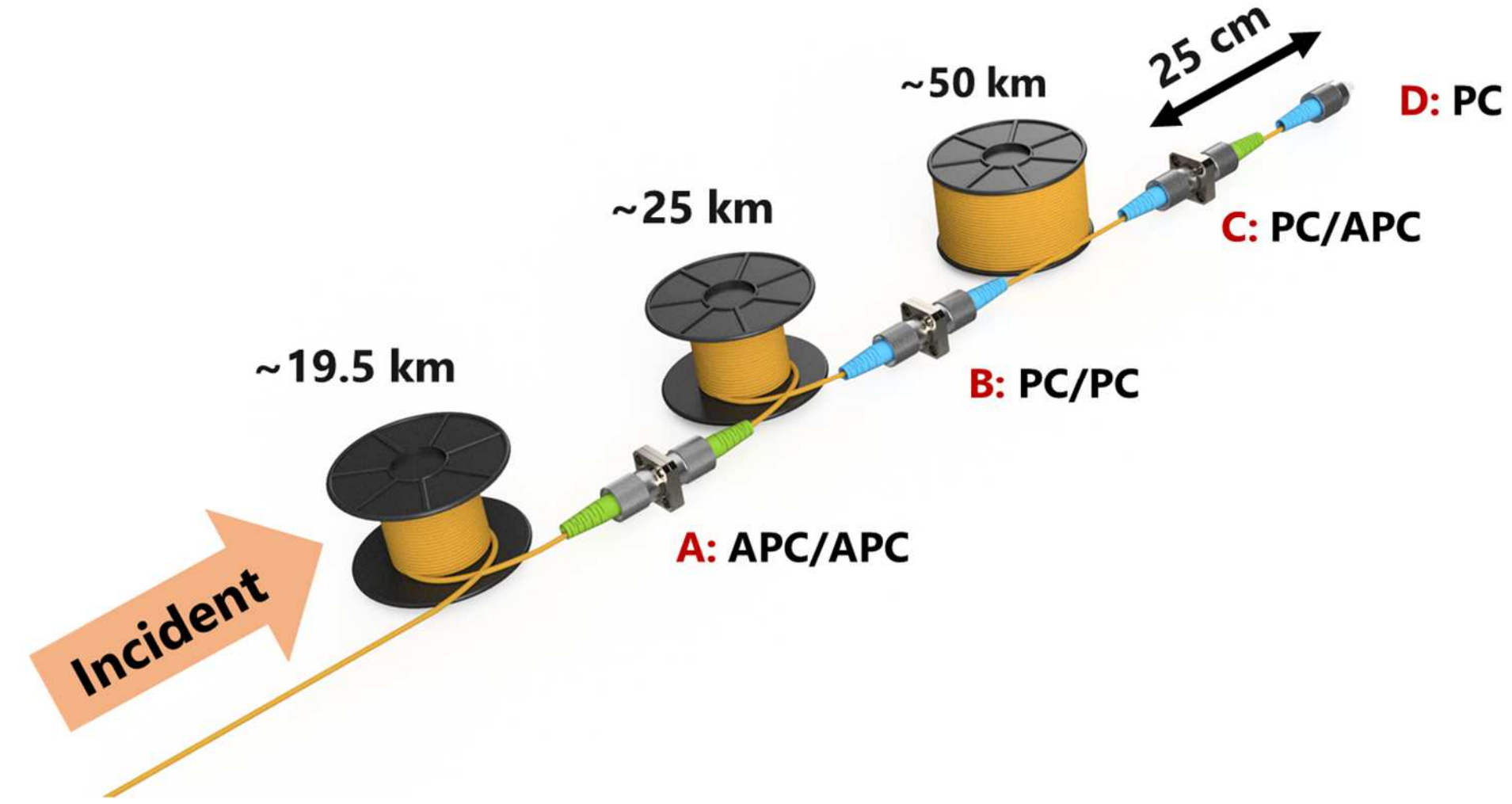

**FIG. 5. Schematic of the fiber under test** (APC: angled physical contact; PC: physical contact). Fiber spools of ~19.5 km, ~25 km, and ~50 km were connected in series to create the connector-induced reflection points A, B, and C. A ~25 cm patch cable was attached at the end to realize reflection points C and D.

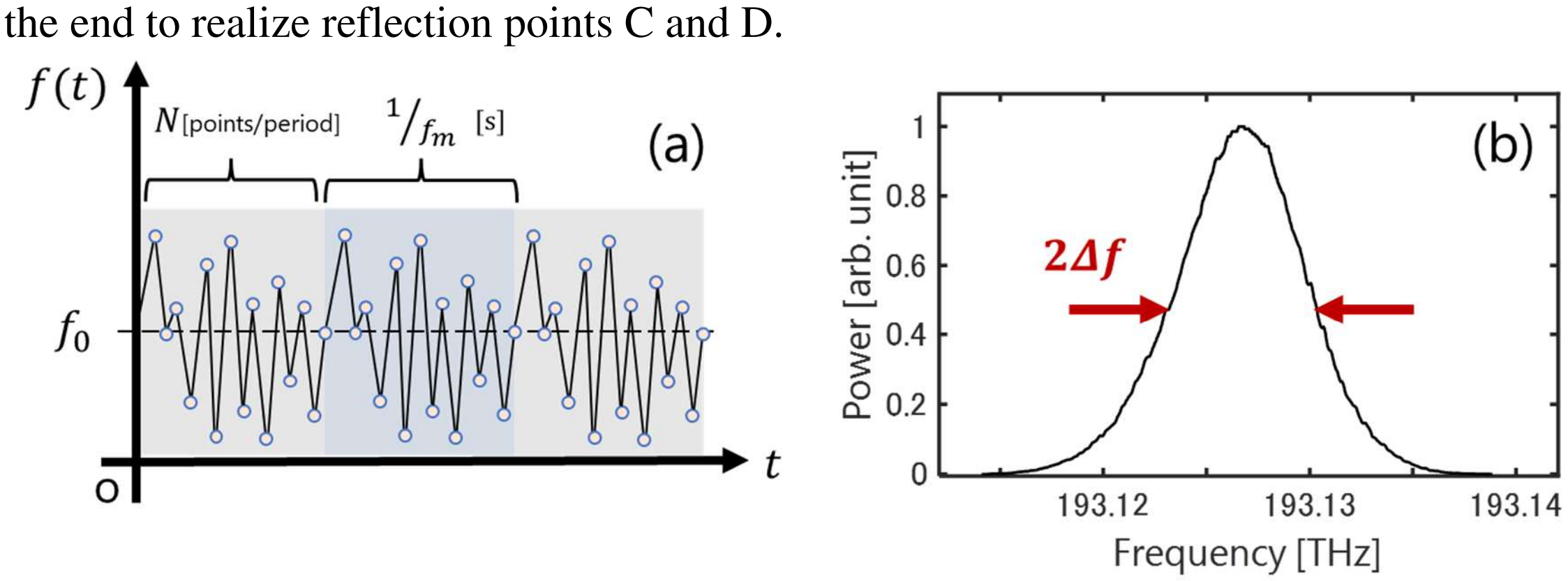

**FIG. 6. Schematic of the periodic pseudo-random modulation (PPRM) and its optical spectrum.** (a) Conceptual time-domain waveform of a PPRM signal. The waveform is represented by $N$ points per cycle, and the cycle period is $1/f_m$. (b) Measured optical spectrum of the light source under PPRM modulation for Fig. 7b. From the FWHM of the spectrum, the modulation amplitude is estimated to be $\Delta f = 3.6$ GHz.

Three erbium-doped fiber amplifiers (EDFAs) were employed to compensate for optical loss in the system. The final amplification stage (EDFA-α) increased the optical power launched into the FUT to 30 dBm. For heterodyne detection, a 40-MHz acousto-optic modulator (AOM) was inserted in the reference arm. The back-reflected light from the FUT (Fig. 5) was amplified to 4.6 dBm using EDFA-γ, while the reference light was amplified to 8.5 dBm using EDFA-β. These two optical signals were combined and detected by a balanced photodiode (BPD).

The electrical output of the BPD was passed through a band-pass filter centered at 40 MHz with a bandwidth of $f_b = 3$ MHz to extract the heterodyne beat component. The signal was subsequently low-pass filtered at $f_v =$ 30 Hz for noise suppression and digitized by an analog-to-digital converter (ADC) operating at a sampling rate of 12.5 kS/s. The digitized data were transferred to a computer, where additional noise reduction was performed using a 20-point moving average before recording.

**Measurement procedure**

To identify the reflector positions along the entire FUT, sinusoidal frequency modulation given by Eq. (2) was first applied to the incident light. The modulation amplitude was set to $\Delta f$ = 15 GHz, and the modulation frequency $f_m$ was swept from 1050 Hz to 300 Hz. This sweep scanned the correlation peak over the entire fiber length according to Eq. (5), with a total scan time of 10 s.

Next, random-access interrogation confined to the vicinity of the fiber end was performed using the PPRM waveform shown in Fig. 6a. In this experiment, the PPRM waveform was generated from a periodically repeated pseudo-random sequence with a Gaussian distribution. The discrete sequence length was $N$ = 32768 points per period. The measured optical spectrum under PPRM exhibited an approximately Gaussian profile with an FWHM of 3.6 GHz, as shown in Fig. 6b. By sweeping the modulation frequency $f_m$ from 520.290 Hz to 520.282 Hz, the correlation peak was scanned locally near the end of the FUT. The scan time for this localized interrogation was 10 s.

# Results

Figures 7a and 7b present the experimental results obtained with sinusoidal modulation and with PPRM, respectively. Both traces are rescaled by min–max normalization so that the minimum and maximum values are set to 0 and 1, respectively. For clarity, in Fig. 7b, the distance axis is redefined so that reflection point C corresponds to 0 m.

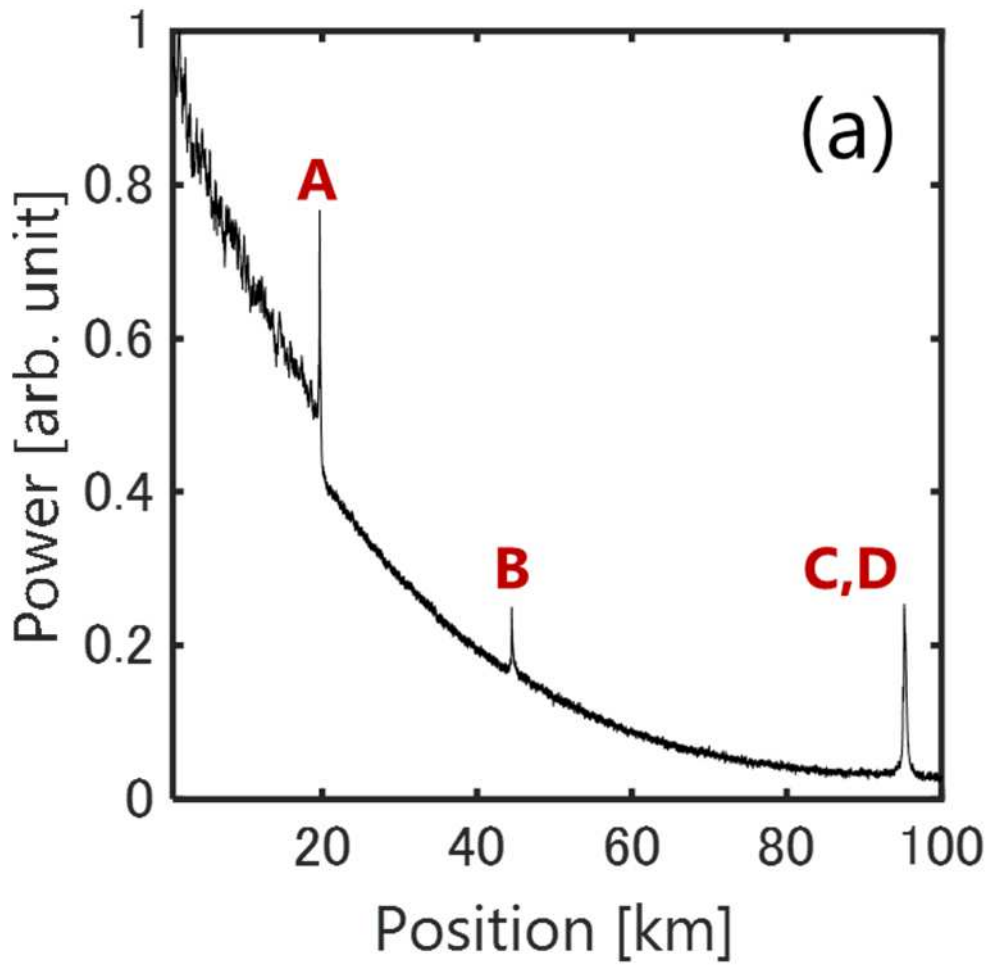

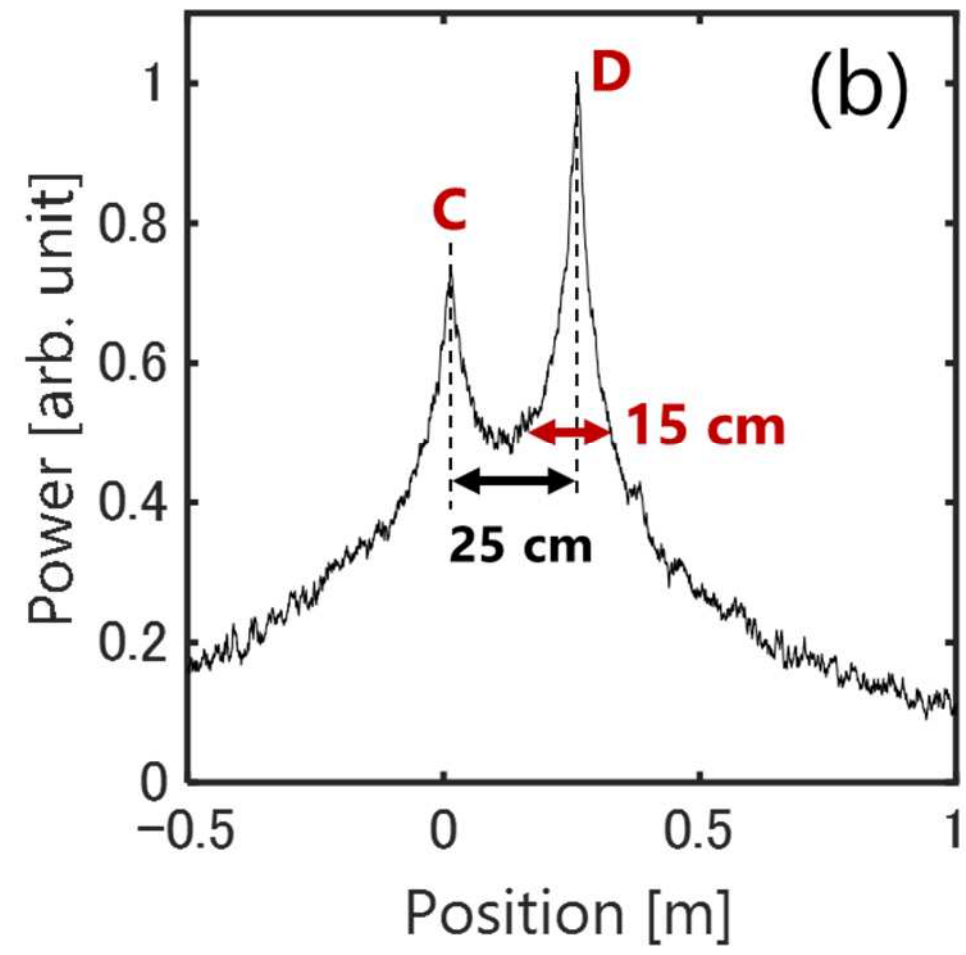

**FIG. 7. Experimental results.** All traces are rescaled using min–max normalization so that the minimum and maximum values are set to 0 and 1, respectively. For Fig. 7b, the distance axis is redefined so that reflection point C corresponds to 0 m. (a) Distributed measurement using conventional sinusoidal frequency modulation. (b) Random-access measurement using PPRM modulation.

Figure 7a shows the distributed measurement result obtained using conventional sinusoidal frequency modulation. Although the overall reflectivity distribution along the FUT can be observed, the reflection points C and D near the fiber end are not clearly resolved. This behavior is consistent with the meter-scale spatial resolution imposed by the range–resolution trade-off inherent to sinusoidal-modulation OCDR under long measurement-range conditions.

In contrast, Fig. 7b shows the random-access measurement result obtained using PPRM modulation. The correlation peak is positioned near the fiber end, enabling localized interrogation without scanning the entire FUT. In this case, the closely spaced reflection points C and D are clearly resolved, with an FWHM of approximately 15 cm for the individual reflection peaks at a distance of approximately 94.5 km. This result confirms that centimeter-order spatial resolution can be achieved even at long distances when the modulation waveform is designed according to the beat-spectrum synthesis principle. The PPRM measurement therefore resolves the two reflection points near the fiber end that are unresolved in the sinusoidal-modulation measurement.

## Discussion

This study shows that the range–resolution coupling of conventional sinusoidal-modulation OCDR can be reduced by controlling the beat spectrum through periodic frequency modulation. We obtained two main results. First, the beat spectrum for arbitrary periodic modulation can be expressed as successive convolutions of the contributions from its harmonic components, linking the modulation waveform to the spatial response. Second, PPRM enabled random-access interrogation near the end of an approximately 100-km fiber and resolved closely spaced reflectors that were not resolved with sinusoidal modulation. These results support beat-spectrum design as a means of combining long-range surveying with localized high-resolution interrogation.

Equation (12) shows that the beat spectrum under periodic modulation is determined by successive convolution of the spectral contributions from its harmonic components. For the waveforms examined in Fig. 3, increasing the harmonic content redistributes the beat-spectrum components over a wider frequency range as the position moves away from the correlation point, reducing the components that remain within the detection bandwidth. This relation connects the harmonic composition of the modulation waveform to the resulting spatial response.

PPRM was used to test this relation experimentally. The sinusoidal-modulation measurement provided a full-length reflectivity trace, but the closely spaced reflection points near the fiber end were not resolved. In contrast, PPRM allowed the same region to be interrogated by random access and separated the two reflection points. The narrower spatial response agrees with the redistribution of the beat spectrum predicted by Eq. (12) and observed in the simulations in Fig. 3.

For fiber-network monitoring, the present method allows full-length surveying and local high-resolution interrogation to be used separately. A sinusoidal-modulation scan can first locate reflection or loss events along the fiber, after which PPRM can be used for random-access interrogation of the region of interest. This two-step operation avoids the need to scan the entire link at high spatial resolution and is suited to locating faults within components such as closures and connector assemblies.

Long-range, high-resolution operation has also been demonstrated using other reflectometry techniques. Chaotic correlation OTDR achieved a spatial resolution of 8.2 cm over 100 km[28], while time-gated digital OFDR achieved 1.6-m resolution over 110 km[29]. The present OCDR method differs in its ability to combine full-length surveying with random-access local interrogation while using a 12.5-kS/s ADC after analog filtering. These features provide a different operating mode rather than a direct improvement in range–resolution performance over OTDR or OFDR.

Several limitations remain. The present setup requires an approximately 200-km optical delay to adjust the relative optical path length, which increases the system footprint. Long-range operation also relies on optical amplification and a high launch power, making noise, dynamic range, and nonlinear effects relevant to practical implementation. In addition, PPRM represents only one waveform within the modulation space described by Eq. (12); optimization of the harmonic amplitudes and phases may further improve the balance among spatial resolution, sidelobe suppression, measurement range, and acquisition time.

# Conclusion

We formulated OCDR under arbitrary periodic frequency modulation by expressing the beat spectrum as successive convolutions of the contributions from harmonic components. Using PPRM, we measured the full-length reflectivity distribution along an approximately 100-km fiber and performed random-access interrogation near the fiber end, where closely spaced reflectors were resolved with a spatial resolution of approximately 15 cm. These results show that beat-spectrum design can reduce the range–resolution coupling of conventional sinusoidal-modulation OCDR and combine long-range surveying with localized high-resolution interrogation.

# Supplementary material

See the supplementary material for additional details on the ghost peaks originating from the spectral feet of the beat spectrum (S1).

# Acknowledgments

This work was partially supported by JSPS KAKENHI (Grant Numbers 21H04555, 24KJ0908, and 26H02136), JST ACT-X (Grant Number JPMJAX25M9), and research grants from the Telecommunications Advancement Foundation and the Asahipen Hikari Foundation.

# Author declarations

### Conflict of interest

The authors declare no competing interests.

### Ethics approval

Ethics approval is not required.

### Author contributions

T.K. and Y.M. conceived and designed the study. T.K., Y.H., and Soshi Yoshida built the experimental setup and performed the measurements. T.K. carried out the numerical simulations and data processing. T.K. and K.M. developed the software used for instrument control and data acquisition. T.K., Sze Yun Set, Shinji Yamashita, and Y.M. analyzed and interpreted the results. Sze Yun Set, Shinji Yamashita, and Y.M. supervised the project. T.K. drafted the manuscript, and all authors reviewed and revised the manuscript.

## Data availability

All data necessary to evaluate the conclusions of this study are included in the paper. Additional data related to this work are available from the corresponding author upon reasonable request.